%% file: IEEE-conference-template-062824.tex
\documentclass[10pt,conference]{IEEEtran}

\IEEEoverridecommandlockouts

\usepackage{cite}
\usepackage{amsmath,amssymb,amsfonts}
\usepackage{algorithmic}
\usepackage{graphicx}
\usepackage{textcomp}
\usepackage{xcolor}
\usepackage{siunitx}

\usepackage[switch]{lineno}

\input{config}

\makeatletter
\DeclareRobustCommand{\change}{%
  \@bsphack
  \leavevmode
  \color{blue}
  \@esphack
}
\DeclareRobustCommand{\stopchange}{%
  \@bsphack
  \normalcolor
  \@esphack
}
\makeatother

\def\BibTeX{{\rm B\kern-.05em{\sc i\kern-.025em b}\kern-.08em
    T\kern-.1667em\lower.7ex\hbox{E}\kern-.125emX}}
\begin{document}

\title{\sysname: Is Your "Clean" Vulnerability Dataset Really Solvable? \\ Exposing and Trapping Undecidable Patches and Beyond}

\makeatletter
\newcommand{\linebreakand}{%
  \end{@IEEEauthorhalign}%
  \hfill\mbox{}\par%
  \mbox{}\hfill\begin{@IEEEauthorhalign}%
}
\makeatother

\author{%
  \IEEEauthorblockN{Zeyu Gao\IEEEauthorrefmark{1}\thanks{\IEEEauthorrefmark{1}Equal contribution}}
  \IEEEauthorblockA{%
    \textit{Tsinghua University}\\
    gaozy22@mails.tsinghua.edu.cn%
  }
  \and
  \IEEEauthorblockN{Junlin Zhou\IEEEauthorrefmark{1}}
  \IEEEauthorblockA{%
    \textit{Sichuan University}\\
    zhoujunlin@stu.scu.edu.cn%
  }
  \linebreakand
  \IEEEauthorblockN{Bolun Zhang}
  \IEEEauthorblockA{%
    \textit{Institute of Information Engineering,}\\
    \textit{Chinese Academy of Sciences}\\
    zhangbolun@iie.ac.cn%
  }
  \and
  \IEEEauthorblockN{Yi He}
  \IEEEauthorblockA{%
    \textit{Wuhan University}\\
    heyi-es@whu.edu.cn%
  }
  \and
  \IEEEauthorblockN{Chao Zhang\IEEEauthorrefmark{2}\thanks{\IEEEauthorrefmark{2}Corresponding author}}
  \IEEEauthorblockA{%
    \textit{Tsinghua University}\\
    chaoz@tsinghua.edu.cn%
  }
  \linebreakand
  \IEEEauthorblockN{Yuxin Cui}
  \IEEEauthorblockA{%
    \textit{Tsinghua University}\\
    yx-cui24@mails.tsinghua.edu.cn%
  }
  \and
  \IEEEauthorblockN{Hao Wang}
  \IEEEauthorblockA{%
    \textit{Tsinghua University}\\
    hao-wang20@mails.tsinghua.edu.cn%
  }
}

\maketitle

\input{abstract}


\input{body}

\bibliographystyle{IEEEtran}
\bibliography{elsa,extra}

\end{document}

%% file: config.tex
\usepackage{verbatim}
\usepackage{amsmath}

\usepackage{amssymb}
\usepackage{algorithmic}
\usepackage{algorithm}
\usepackage{makecell}
\usepackage{boxedminipage}
\usepackage{svg}
\usepackage{amsmath}
\definecolor{shadecolor}{RGB}{246,246,246}

\usepackage{graphicx}
\usepackage{epstopdf}
\usepackage{array}
\usepackage{booktabs}
\usepackage{multirow}
\usepackage{colortbl}
\usepackage{caption} 
\usepackage{multicol}
\usepackage{tabularx}
\usepackage{framed}
\usepackage{enumitem}
\usepackage{threeparttable}
\usepackage{supertabular}
\graphicspath{{figs/}}
\usepackage{xcolor}
\usepackage{colortbl} 
\definecolor{jade}{rgb}{0.0, 0.66, 0.42}
\definecolor{carolinablue}{rgb}{0.6, 0.73, 0.89}
\definecolor{dkgreen}{rgb}{0,0.6,0}
\definecolor{dkblue}{rgb}{0,0.4,0.5}
\definecolor{gray}{rgb}{0.5,0.5,0.5}
\definecolor{mauve}{rgb}{0.58,0,0.82}

\definecolor{codegreen}{rgb}{0,0.6,0}
\definecolor{codegray}{rgb}{0.5,0.5,0.5}
\definecolor{codepurple}{rgb}{0.58,0,0.82}
\definecolor{codeblue}{rgb}{0,0,205}
\definecolor{backcolour}{rgb}{245,245,245}
\usepackage{listings}
\usepackage{enumitem}

\lstdefinelanguage
   [x64]{Assembler}     
   [x86masm]{Assembler} 
   {morekeywords={xend, CDQE,CQO,CMPSQ,CMPXCHG16B,JRCXZ,LODSQ,MOVSXD, %
                  POPFQ,PUSHFQ,SCASQ,STOSQ,IRETQ,RDTSCP,SWAPGS, %
                  rax,rdx,rcx,rbx,rsi,rdi,rsp,rbp, %
                  r8,r8d,r8w,r8b,r9,r9d,r9w,r9b, %
                  r10,r10d,r10w,r10b,r11,r11d,r11w,r11b, %
                  r12,r12d,r12w,r12b,r13,r13d,r13w,r13b, %
                  r14,r14d,r14w,r14b,r15,r15d,r15w,r15b}} 

\lstdefinestyle{mystyle}{
    backgroundcolor=\color{backcolour},   
    commentstyle=\color{codegreen},
    keywordstyle=\color{codeblue},
    numberstyle=\tiny\color{codegray},
    stringstyle=\color{codeblue},
    basicstyle=\ttfamily\footnotesize,
    breakatwhitespace=false,         
    breaklines=true,                 
    captionpos=b,                    
    keepspaces=true,                 
    numbers=left,                    
    numbersep=5pt,                  
    showspaces=false,                
    showstringspaces=false,
    showtabs=false,                  
    tabsize=2
}
\PassOptionsToPackage{hyphens}{url}\usepackage{hyperref}
\usepackage{url}
\usepackage{flushend}
\usepackage{balance}

\usepackage[misc]{ifsym}
\usepackage{bbding}
\usepackage{xspace}

\newcommand{\sysname}{\textsc{mono}\xspace}
\newcommand{\dsname}{\textsc{MonoLens}\xspace}

\usepackage{xspace}
\usepackage{subfig}

\usepackage[T1]{fontenc}

\usepackage{tabularx}
\usepackage{ragged2e}
\usepackage{booktabs}
\usepackage{fontawesome5}

\renewcommand{\arraystretch}{1.2}
\newcommand{\cmark}{\textcolor{green!70!black}{\faCheckCircle}}
\newcommand{\tmark}{\textcolor{orange!80!black}{\faAdjust}}
\newcommand{\xmark}{\textcolor{red!70!black}{\faTimesCircle}}
 \usepackage[markup=underlined]{changes}

\usepackage{multirow}    
\usepackage{amssymb}     
\usepackage{booktabs}    

\usepackage{balance}

\usepackage{tcolorbox}

\newcounter{answerctr}
\newcommand{\answer}[1]{
    \begin{tcolorbox}[
        colback=gray!5, 
        left=1pt, 
        right=1pt, 
        top=2pt, 
        bottom=2pt,
        boxrule=0.75pt,
        before skip=2pt,
        after skip=2pt
    ]
    \textbf{Answering RQ\arabic{answerctr}:} #1 
    \end{tcolorbox}
    \refstepcounter{answerctr} 
}
\usepackage{array,tabularx,booktabs,threeparttable}
\usepackage{siunitx}
\renewcommand{\arraystretch}{0.95}

\newcolumntype{Y}{>{\raggedleft\arraybackslash}X}
\newcolumntype{C}{>{\centering\arraybackslash}X}                 
\newcolumntype{N}{S[table-format=4]}                             
\newcolumntype{P}{S[
    table-format=3.2,
    table-space-text-post=\%, 
    table-space-text-pre=\phantom{0}
  ]}
\usepackage{adjustbox}

%% file: abstract.tex
\begin{abstract}

The quantity and quality of vulnerability datasets are essential for developing deep learning solutions to vulnerability-related tasks. Due to the limited availability of vulnerabilities, a common approach to building such datasets is analyzing security patches in source code. However, existing security patches often suffer from inaccurate labels, insufficient contextual information, and undecidable patches that fail to clearly represent the root causes of vulnerabilities or their fixes. 
These issues introduce noise into the dataset, which can mislead detection models and undermine their effectiveness.
To address these issues, we present \sysname, a novel LLM-powered framework that simulates human experts' reasoning process to construct reliable vulnerability datasets. 
\sysname introduces three key components to improve security patch datasets: (i) semantic-aware patch classification for precise vulnerability labeling, (ii) iterative contextual analysis for comprehensive code understanding, and (iii) systematic root cause analysis to identify and filter undecidable patches. 
Our comprehensive evaluation on the MegaVul benchmark demonstrates that \sysname can correct 31.0\% of labeling errors, recover 89\% of inter-procedural vulnerabilities, and reveals that 16.7\% of CVEs contain undecidable patches. 
Furthermore, \sysname's enriched context representation improves existing models' vulnerability detection accuracy by 15\%.
\end{abstract}

%% file: body.tex
\section{Introduction}
The number of exposed software vulnerabilities significantly grow in recent years, drawing attentions to many vulnerability-related tasks, e.g., vulnerability discovery.
As popular as fuzzing~\cite{fioraldiAFLCombiningIncremental2020, fioraldiLibAFLFrameworkBuild2022,GoogleSyzkaller2015}, which is the most popular solution to finding vulnerabilities\footnote{
For instance, platforms like OSS-Fuzz~\cite{OSSFuzz} and syzbot~\cite{Syzbot} collectively identifying over 10,000 vulnerabilities in critical open-source projects.},
deep learning (DL) based approaches~\cite{fuLineVulTransformerbasedLineLevel2022,hanifVulBERTaSimplifiedSource2022,wuVulCNNImageinspiredScalable2022,zhouDevignEffectiveVulnerability2019,chakrabortyDeepLearningBased2020,liVulDeePeckerDeepLearningBased2018}
quickly emerge and have shown promising results.
They have achieved state-of-the-art performance (90\% accuracy) on curated datasets like Devign~\cite{zhouDevignEffectiveVulnerability2019} and BigVul~\cite{fanBigVulCodeVulnerabilityDataset2020} by utilizing advanced neural architectures including graph neural networks and transformer models.
While Large Language Models (LLMs) are deeply explored in other software engineering domains~\cite{jimenezSWEbenchCanLanguage2023,yangSWEagentAgentComputerInterfaces2024,xiaAgentlessDemystifyingLLMbased2024},
they are also leveraged to detect vulnerabilities~\cite{ullahLLMsCannotReliably2023,zibaeiradVulnLLMEvalFrameworkEvaluating2024,dingVulnerabilityDetectionCode2024,liuVulDetectBenchEvaluatingDeep2024,khareUnderstandingEffectivenessLarge2024}.

However, practical deployments of DL-based approaches (including LLMs)
are facing persistent challenges in real-world applications due to training dataset inaccuracy~\cite{croftDataQualitySoftware2023,guoDataQualityIssues2023,liCleanVulAutomaticFunctionLevel2024,dingVulnerabilityDetectionCode2024}, limited model interpretability~\cite{huInterpretersGNNBasedVulnerability2023,risseTopScoreWrong2024}, and heavy dependency on model architecture~\cite{fuLineVulTransformerbasedLineLevel2022}. 
Specifically, for vulnerability related tasks, the quantity and quality of training dataset are both limited.  
Due to the limited availability of vulnerabilities, a common approach to building such datasets is analyzing security patches in source code. 
However, existing security patches are noisy.

One primary manifestation of noise is \textit{semantic mislabeling}, which occurs when security patches are conflated with non-security-related patches like feature updates or general bug fixes
~\cite{wangReposVulRepositoryLevelHighQuality2024,dingVulnerabilityDetectionCode2024,nieUnderstandingTacklingLabel2023}. 
Recent researches~\cite{liCleanVulAutomaticFunctionLevel2024,croftDataQualitySoftware2023,guoDataQualityIssues2023} show that 
this is common.
The widely used datasets like BigVul~\cite{fanBigVulCodeVulnerabilityDataset2020} and DiverseVul~\cite{chenDiverseVulNewVulnerable2023} exhibit label accuracy rates below 60\% as reported by~\cite{liCleanVulAutomaticFunctionLevel2024}. 
But due to \textbf{inadequate handling of label ambiguity (L1)}, current methods~\cite{liCleanVulAutomaticFunctionLevel2024,wangReposVulRepositoryLevelHighQuality2024,dingVulnerabilityDetectionCode2024}, employing heuristic rules or basic LLM prompts to filter the non-security-related patches, frequently misclassify patches because accurate labeling necessitates a holistic understanding derived from patches, commit messages, and auxiliary information. 

Another key manifestation of noise is \textit{inter-procedure ambiguity}, which arises when vulnerability triggers depend on multi-function interactions~\cite{wenRepositoryLevelGraphRepresentation,liEverythingYouWanted2025}. 
For example, analyzing a function in isolation is difficult to determine if an input pointer could cause a null pointer dereference. Such a determination depends on whether the caller might pass a null value.
Bcause of \textbf{reliance on rule-based context extraction (L2)}, 
previous methods either extract the entire repository call graph to preserve as much context as possible~\cite{wangReposVulRepositoryLevelHighQuality2024}, which leads to information explosion with overwhelming irrelevant data~\cite{anWhyDoesEffective2024}, or extract a fixed number of callee layers~\cite{liEverythingYouWanted2025}, which results in incomplete context for some vulnerabilities, making them undetectable and introducing noise.

More critically, we identify \textit{undecidable patches}, a challenging new category of noisy patches that often address subtle vulnerabilities, such as logic errors causing memory corruption or denial-of-service (DoS) due to improper boundary conditions. 
But these implicit vulnerabilities always lack obvious vulnerability indicators or clear links between the patch behavior and CWE taxonomies.
Thus, the presence of these vulnerabilities in the original code is often difficult to ascertain using static analysis solutions alone, and only become evident after reviewing the patch and associated discussions.
To better illustrate their nature, prevalence and diversity of sources, we present concrete examples in Section~\ref{sec:undecidable_flaws}. The unawareness of them leads to a \textbf{lack of mechanisms to handle undecidable patches (L3)} in current methods.
This deficiency means vulnerabilities, hard for static analysis to detect, persist in existing datasets. Training models on such ambiguous instances can lead to them learning spurious or non-existent vulnerability patterns, causing hallucinations model.

\textbf{Our Solution.} To address the limitations of prior approaches in handling these types of noise, we propose \sysname (\textbf{M}ulti-agent \textbf{O}perated \textbf{N}oise \textbf{O}utfilter), an LLM-powered framework that emulates human expert analysis to tackle data quality challenges in vulnerability dataset construction. 
Our approach introduces three key innovative solutions:

\begin{itemize}  
    \item \textbf{S1: Semantic-Aware Patch Classification.} \sysname leverages the natural language understanding and code analysis capabilities of LLMs to evaluate commit messages, patches, and auxiliary information such as pull requests. 
    This enables a nuanced classification of patches and accurately distinguishes security-related fixes from non-security patches.

   \item \textbf{S2: Iterative Multi-Agent Contextual Analysis.} \sysname employs a multi-agent architecture that combines static analysis tools with LLM reasoning. Through an iterative process, the agents progressively gather relevant code context from the repository, enabling in-depth reasoning to capture the intricacies of each potential vulnerability.

    \item \textbf{S3: Vulnerability Root Cause Analysis and Undecidable Patch Filtering.} \sysname performs in-depth vulnerability root cause analysis by tracing execution paths and data flows within the project. This process identifies the root cause of vulnerabilities and filters out \textit{undecidable patches}, where the root cause cannot be determined from the available static context.
\end{itemize}

Through a comprehensive evaluation on the MegaVul vulnerability dataset~\cite{niMegaVulVulnerabilityDataset2024}, we demonstrate the effectiveness of \sysname. It identifies that over 30\% of patches in MegaVul are non-security patches and reveals residual noise even in existing cleaned datasets. 
To analyze patch root causes, \sysname extracts an average of 3.43 contextual code snippets, with 89\% of patches requiring more than one piece of context. For cases where \sysname cannot determine a root cause, over 80\% are classified as undecidable patches. We estimate that approximately 16.7\% of the MegaVul dataset consists of such undecidable patches, representing a significant portion.

Our contributions are as follows:

\begin{itemize}
    \item We identify and formalize a new but challenging category of label noise in the current vulnerability dataset, termed \textit{Undecidable patches}. We estimate that approximately 16.7\% of the MegaVul dataset falls into this category.
    \item We propose the first end-to-end LLM-powered dataset construction framework that implements expert reasoning patterns through multi-agent collaboration, leveraging static analysis tools to produce a high-quality dataset.
    \item Our empirical validation demonstrates a 15\% improvement in downstream vulnerability detection model performance when using the context provided by \sysname.
    \item We open source the framework \sysname and the dataset \dsname in \href{https://github.com/vul337/mono}{https://github.com/vul337/mono} to facilitate future research.
\end{itemize}

\input{comparison-table.tex}

\section{Background and Motivation}

\subsection{Noisy Patches that Hinder Model Training}

As shown in Table~\ref{tab:full-comparison}, previous studies~\cite{liCleanVulAutomaticFunctionLevel2024,risseTopScoreWrong2024,wangReposVulRepositoryLevelHighQuality2024,dingVulnerabilityDetectionCode2024} focus on two critical noise issues in prevailing vulnerability datasets that degrade model performance. 
The first issue, termed \textit{semantic mislabeling}, arises when non-security-related patches (e.g., testing improvements, feature updates, or general bug fixes) are mistakenly labeled as vulnerabilities. 
The second issue, termed \textit{inter-procedure ambiguity}, stems from the lack of multi-function context in most datasets, which typically only provide function-level views of the patch codes.
Training on such datasets often leads to false positives, such as misclassifying functional updates as vulnerabilities or wrongly flagging callers for not validating callee returns despite the callee ensuring its safety.
These issues significantly compromise the reliability of vulnerability detection models.

\subsection{A New Kind of Noise Patch: Undecidable Patches}\label{sec:undecidable_flaws}

In addition to the two aforementioned types of noisy patches, we identify a new category, termed \textit{undecidable patches}. 
These patches fix real vulnerabilities but are extremely difficult to identify in the original code using static analysis. 
In many cases, it is nearly impossible to recognize these patches as security fixes using manual rules. 
For human experts, the vulnerability often only becomes apparent after reviewing the patch—an ``aha'' moment. 
Before the fix is applied, the vulnerabilities remain implicit, with no obvious exploit patterns or clear violations of coding conventions.

These undecidable patches pose a fundamental challenge to automated vulnerability detection. Training models on these ambiguous examples is counterproductive. When a ``vulnerable'' label has no clear evidence in the code, models can learn spurious correlations or ``hallucinate'' patterns that do not exist, 
severely undermining their reliability in real-world scenarios. 
To identify and filter out these patches, we first conduct an empirical study to clearly define these patches. 
Then, we propose the \sysname framework, which purifies vulnerability datasets by removing these harmful undecidable patches.

We identify five common patterns of undecidable patches and illustrate them using real-world examples.

\subsubsection{Involving Runtime Information or High-Level Program Understanding}
\begin{figure}[t!]
  \centering
  \includegraphics[width=0.99\linewidth]{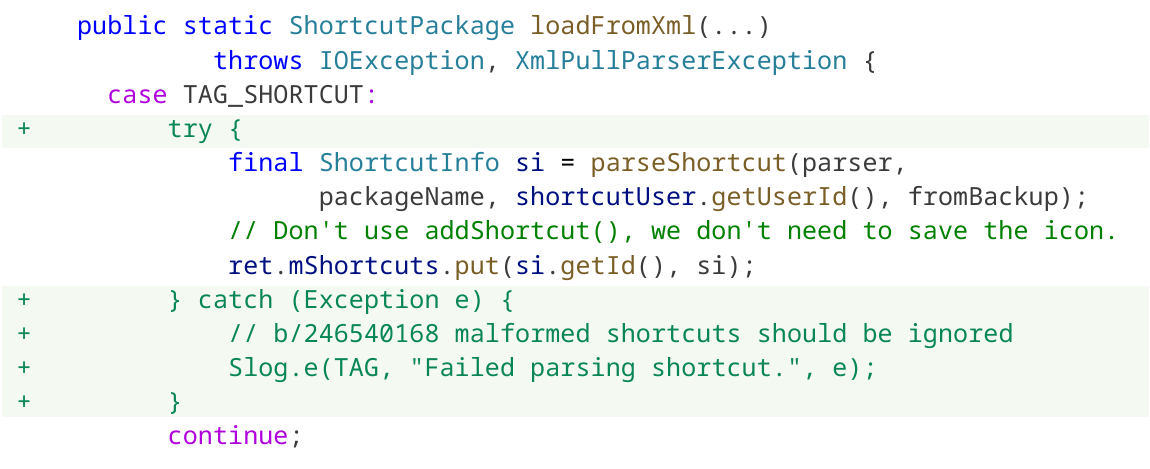}
  \caption{
    \href{https://github.com/PixelExperience/frameworks_base/commit/d5122bfaf18f1503e73c1a3a177a56d0f604a008}{Commit} for fixing \href{https://nvd.nist.gov/vuln/detail/cve-2022-20500}{CVE-2022-20500} by catch an underlying exception to prevent boot-loop.
  }
  \label{fig:cve-2022-20500}
\end{figure}

This kind of patch is difficult to identify with static rules because it relies on runtime information and often involves hidden system-level consequences or unpredictable execution paths.
For instance, consider CVE-2022-20500 (Figure~\ref{fig:cve-2022-20500}), where a patch introduced a \texttt{try-catch} block to prevent a system boot-loop. This fix is not immediately obvious because, from a static perspective, the original code appeared valid. The method already declared \texttt{throws Exception}, placing the responsibility for handling it on the caller, as per standard Java practices. The vulnerability, however, did not stem from the local logic of the code but rather from an unexpected and severe system-level side effect.
In isolation, the original code seems correct. 
However, without runtime context, an automated tool might overfit in such cases and incorrectly apply similar fixes to other safe code.

\subsubsection{Complex Logic-Dependent Issues}
\begin{figure}[t!]
  \centering
  \includegraphics[width=0.95\linewidth]{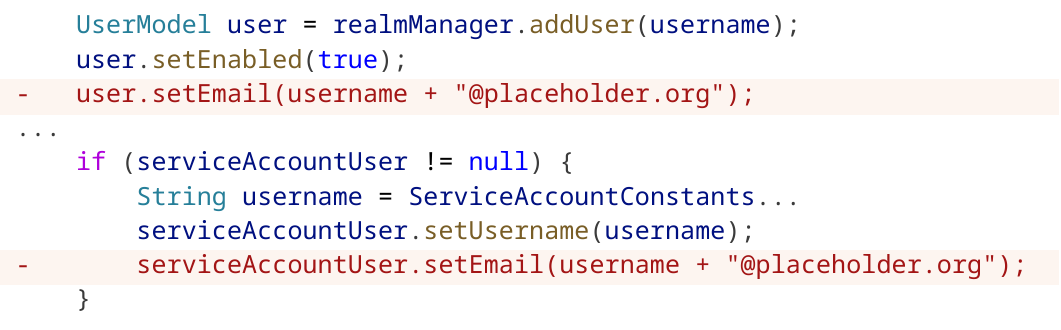}
  \caption{
    \href{https://github.com/keycloak/keycloak/commit/9a7c1a91a59ab85e7f8889a505be04a71580777f}{Commit} for fixing \href{https://nvd.nist.gov/vuln/detail/cve-2019-14837}{CVE-2019-14837} by not creating placeholder e-mails.
  }
  \label{fig:CVE-2019-14837}
\end{figure}

\begin{figure}[t!]
  \centering
  \includegraphics[width=0.77\linewidth]{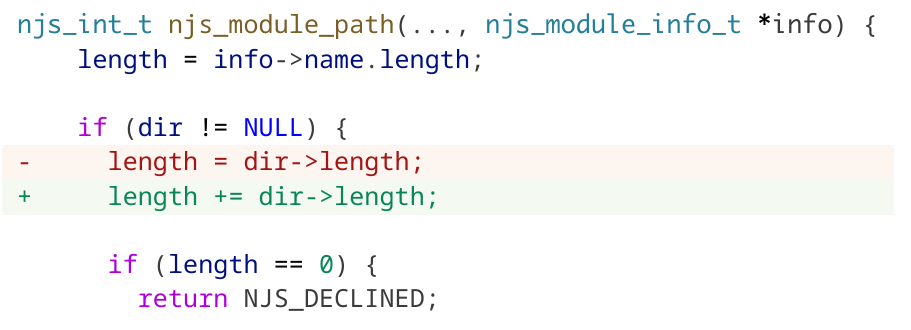}
  \caption{
    \href{https://github.com/nginx/njs/commit/ab1702c7af9959366a5ddc4a75b4357d4e9ebdc1}{Commit} for fixing \href{https://nvd.nist.gov/vuln/detail/CVE-2022-29379}{CVE-2022-29379}. But multiple third parties dispute this report and it is only found in unreleased development code that was not part of the following release.
  }
  \label{fig:CVE-2022-29379}
\end{figure}

While some vulnerabilities are tied to runtime context, another category of undecidable patches is purely internal, arising from violations of an application's complex or unstated logic. In these cases, the code is not syntactically wrong but fails to meet an implicit operational goal, often known only to the developers.
For example, a patch for CVE-2019-14837 (Fig~\ref{fig:CVE-2019-14837}) removed the assignment of a placeholder email to a service account. While this assignment action is syntactically harmless, it violated an unstated rule that such accounts should not have fake emails, preventing potential data integrity failures or errors in other subsystems. 
Similarly, another fix (CVE-2022-29379, Fig~\ref{fig:CVE-2022-29379}) corrected a calculation by changing a single assignment operator. 
In both scenarios,
the original code is perfectly valid from a static analysis viewpoint. The vulnerabilities stem from a deviation from an unstated operational goal—much like violating a formal specification (e.g., an RFC) that was never written down. Without knowing these implicit requirements, an automated tool has no basis for flagging the code as faulty.

\subsubsection{Ambiguous Defensive Programming}
\begin{figure}[t!]
  \centering
  \includegraphics[width=0.80\linewidth]{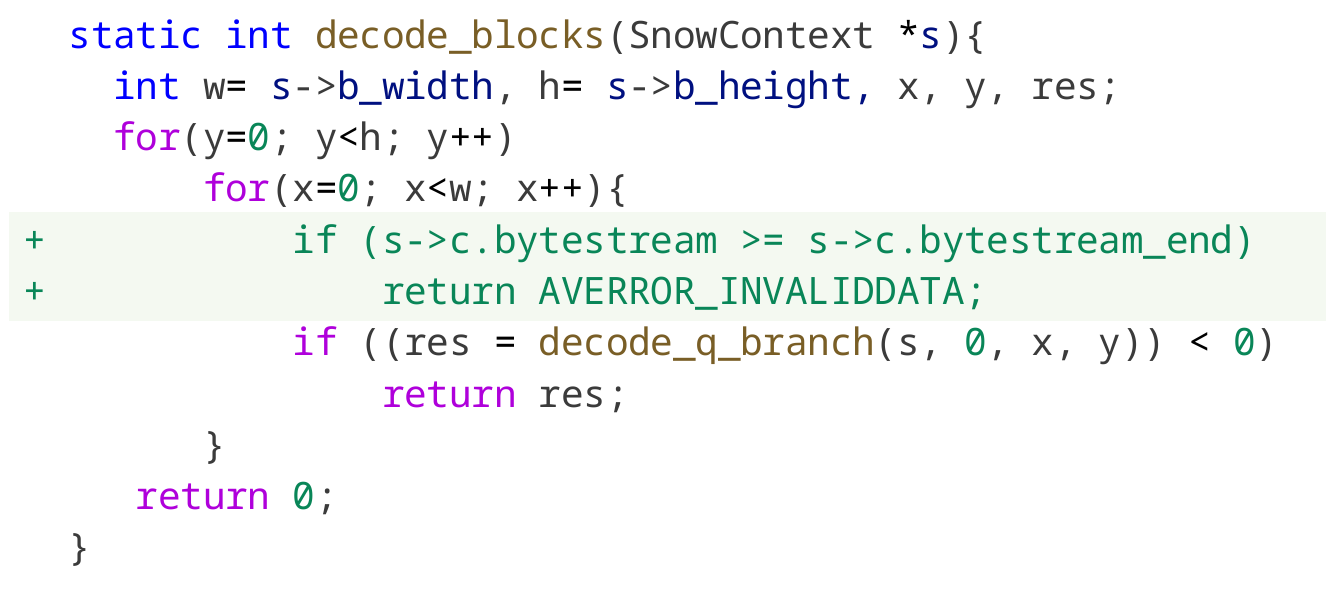}
  \caption{
    \href{https://github.com/FFmpeg/FFmpeg/commit/4527ec2}{Commit} in Devign Dataset, No. 5360. A boundary check is added, but the sink for this check is not clear.
  }
  \label{fig:Devign-5360}
\end{figure}

These are patches that introduce checks which seem like good practice, but whose necessity and specific placement are not evident from the surrounding context.
Devign-5360 (Fig~\ref{fig:Devign-5360}) exemplifies this, where a boundary check
is added inside a loop before calling the \texttt{decode\_q\_branch} function. On the surface, this is a sensible patch to prevent a buffer over-read. The ambiguity, however, lies in two things: first, it is exceptionally difficult to prove statically that callee will actually read out-of-bounds without this check. Second, experts question why the check is needed here in the caller, rather than inside the callee  itself. The patch could be either a critical fix for a hidden vulnerability or simply an overly cautious defense due to uncertainty about the callee's behavior.
Training a model on such an instance might lead it to become overly aggressive, flagging similar patterns as vulnerable.

\subsubsection{Reliance on External Knowledge or Conventions}

\begin{figure}[t!]
  \centering
  \includegraphics[width=0.60\linewidth]{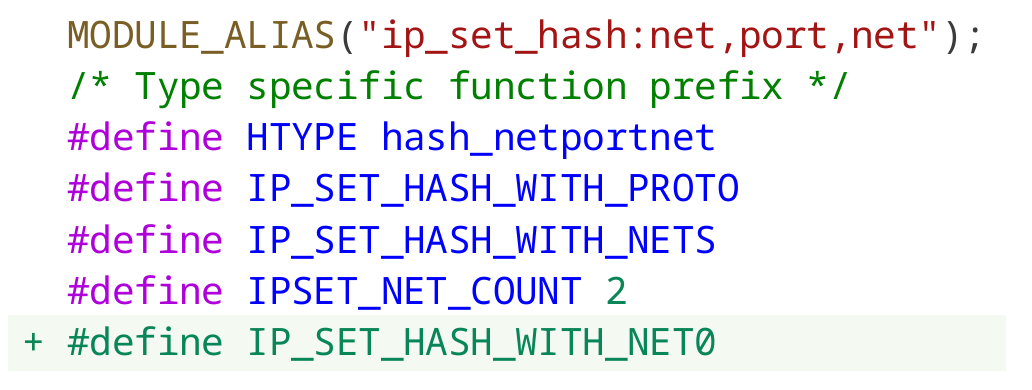}
  \caption{
    Patch for \href{https://nvd.nist.gov/vuln/detail/CVE-2023-42753}{CVE-2023-42753}. Only a marco is added to correct the calucation of the offset of one varaible.
  }
  \label{fig:CVE-2023-42753}
\end{figure}

\begin{figure}[t!]
  \centering
  \includegraphics[width=0.90\linewidth]{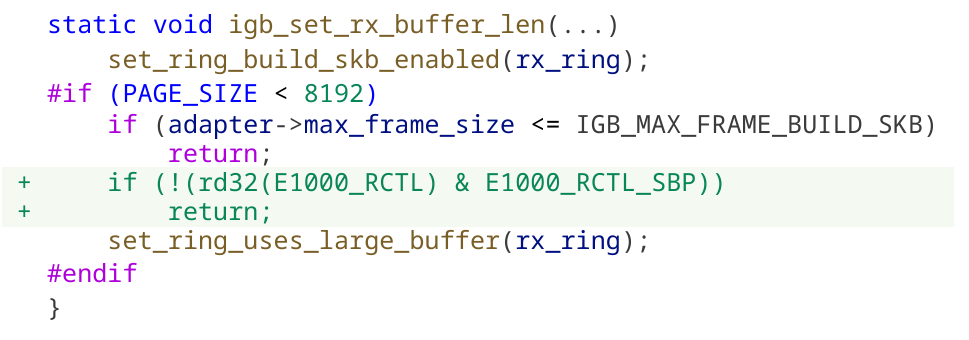}
  \caption{
    A device specific checking is added for \href{https://nvd.nist.gov/vuln/detail/CVE-2023-45871}{CVE-2023-45871}.
  }
  \label{fig:CVE-2023-45871}
\end{figure}

Some fixes address issues that violate external constraints, such as library API usage conventions, security policies, or domain-specific knowledge not present in the source code itself.
In CVE-2023-42753 (Fig~\ref{fig:CVE-2023-42753}), the patch introduces a macro whose necessity depends on external knowledge. However, vulnerability detection in ~\cite{ullahLLMsCannotReliably2023} operates at the file level, ignoring that the macro could be placed elsewhere (e.g., in headers or build configs) and its broader contextual dependencies.
Similarly, CVE-2023-45871 (Fig~\ref{fig:CVE-2023-45871}) relates to a specific SBP bit in a device's data packet, requiring device-specific information for comprehension.
These vulnerabilities cannot be understood by merely inspecting the local code; they require awareness of external specifications or implicit operational contracts.

For such problems, a data-driven approach where models learn from new vulnerability patch data is highly suitable, enabling models to acquire and internalize relevant knowledge. However, the model requires sufficient context to understand the vulnerability's root cause. This ensures it learns genuine vulnerability patterns, not spurious ones as seen in Fig~\ref{fig:CVE-2023-42753}.

\subsubsection{Misclassified Functional Patches or Optimizations}
\begin{figure}[t!]
  \centering
  \includegraphics[width=0.90\linewidth]{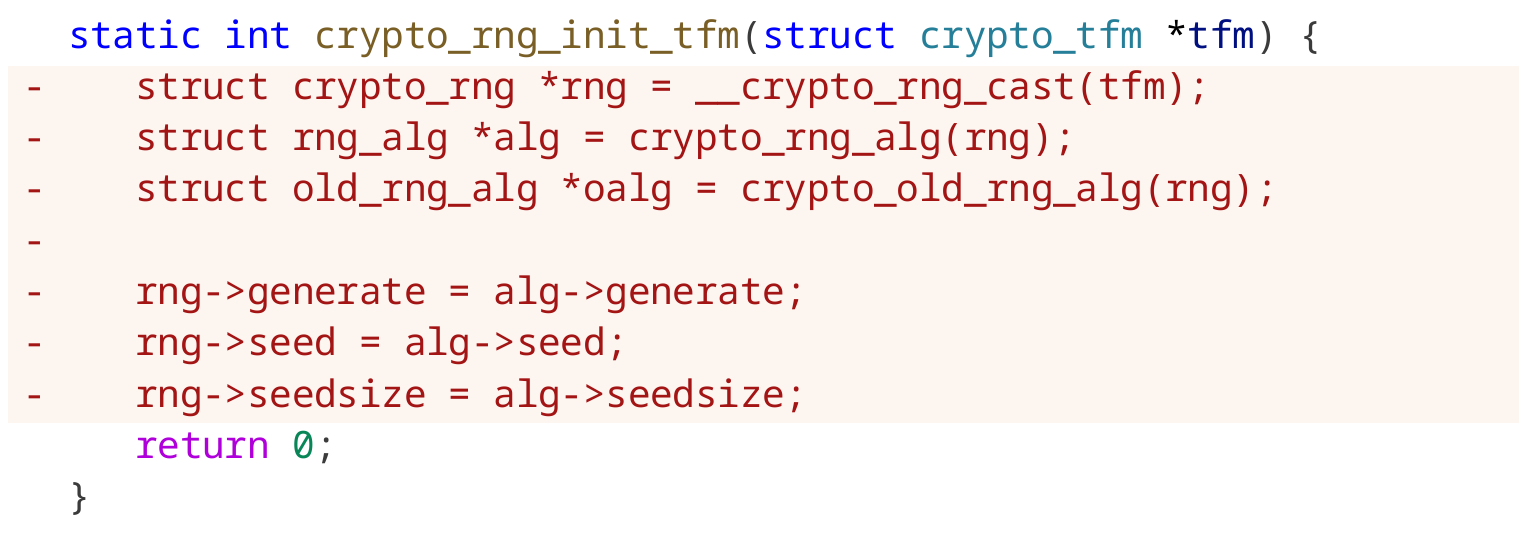}
  \caption{
    Partial patch for fixing \href{https://nvd.nist.gov/vuln/detail/CVE-2017-15116}{CVE-2017-15116}, containing a interface refactoring, removing everything in this function. While the \href{https://git.kernel.org/pub/scm/linux/kernel/git/torvalds/linux.git/commit/?id=94f1bb15bed84ad6c893916b7e7b9db6f1d7eec6}{commit} contains multiple modifications, the CleanVul dataset only retains this diff as part of its final dataset.
  }
  \label{fig:CVE-2017-15116}
\end{figure}

Occasionally, patches that are primarily performance optimizations or feature enhancements are assigned CVEs. In such cases, the ``vulnerability'' and its ``fix'' more closely resemble the addition of new functionality or refinement of existing logic, rather than the remediation of a distinct security flaw.
For example, in CVE-2017-15116 (Fig~\ref{fig:CVE-2017-15116}), involves a refactoring to a new interface, but the patch collected by CleanVul only contains a function cleanup.
Besides, \href{https://github.com/qemu/qemu/commit/3e9fab690d59ac15956c3733fe0794ce1ae4c4af}{Devign-5533} and \href{https://github.com/xen-project/xen/commit/14c3f68a57361f20be33ec3848f83d8636a0d34e}{CVE-2018-10982} are similar cases where the patch is a refactoring of the code, stating ``Those are not deemed to be security issues, but rather
quirks of the current implementation.'' in the commit message.
The root cause of the purported vulnerability is often not evident in the surrounding code context, making the flaw statically undecidable.

Including such undecidable patches in training datasets can be detrimental. Models might learn to associate generic coding patterns (e.g., any exception handling) with vulnerabilities, or they might become overly sensitive to defensive programming constructs without understanding the specific threat they mitigate. By identifying and separating these instances, we aim to create cleaner datasets and highlight areas where current automated analysis techniques fall short, paving the way for more nuanced approaches to vulnerability understanding.

\subsection{Limitations of Existing Patch Identification Approaches}

Previous works fail to effectively handle all these noise categories, as summarized in Table~\ref{tab:comparison-table}. 
In particular, they face significant limitations in addressing semantic mislabeling and inter-procedure ambiguity. This is because they primarily rely on rule-based methods, which lack the nuanced understanding required for comprehensive noise removal. 
Furthermore, prior work neither identifies nor addresses \textit{undecidable patches}, which are estimated to affect approximately 16.7\% of the patches in MegaVul. While manual curation can partially address these issues, it is not scalable for large-scale datasets. 
To address these challenges, we propose \sysname, a framework designed to improve vulnerability datasets by properly handling noisy patches.

\section{Methodology}

Our proposed framework, \sysname, employs a two-stage methodology to enhance vulnerability datasets.
The first stage, focuses on initial patch pre-filtering (Section~\ref{subsec:patch-pre-filtering-classification}), and repository preprocessing (Section~\ref{subsec:data-acquisition-contextual-preprocessing}), where an LLM filters out patches primarily related to non-security-related patches. The second stage, involves an agent attempting to gather sufficient contextual information from the repository to understand the vulnerability's root cause through iterative contextual analysis (Section~\ref{subsec:iterative-contextual-analysis}) and subsequent dataset construction (Section~\ref{subsec:dataset-construction-postprocessing}). If a complete root cause cannot be established with the available static context and the LLM's inherent knowledge, the vulnerability is flagged as potentially undecidable.

\subsection{Patch Pre-filtering and Classification}
\label{subsec:patch-pre-filtering-classification}

\begin{figure*}[t!]
  \centering
  \includegraphics[width=0.75\linewidth]{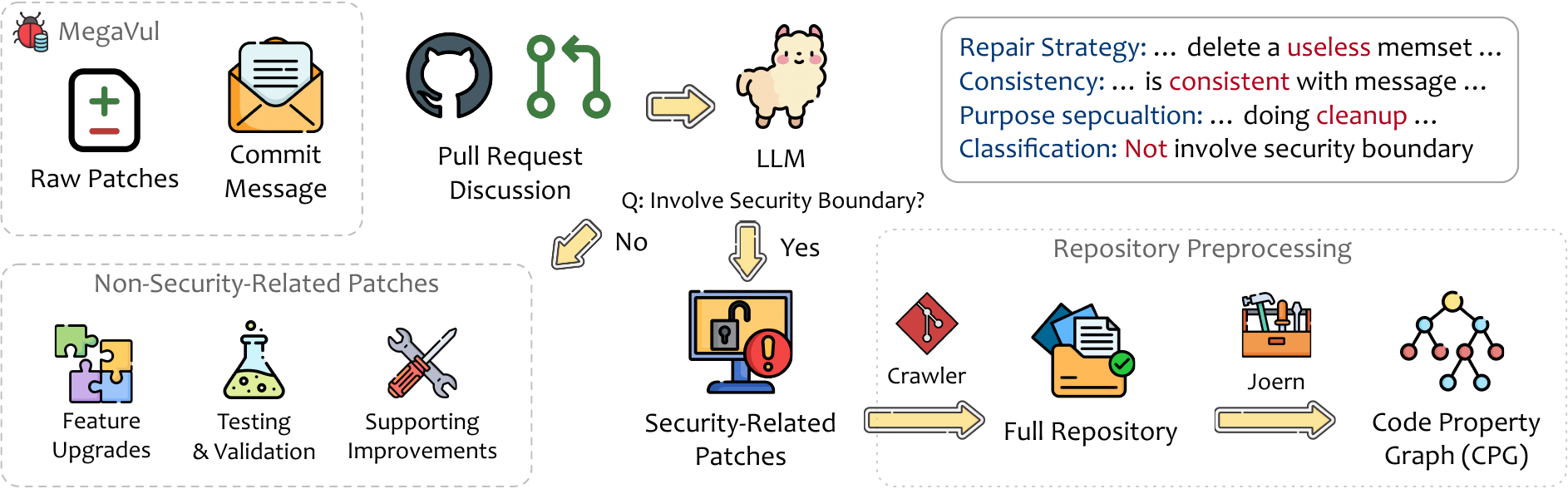}
  \caption{
    Workflow overview of patch pre-filtering and repository preprocessing.
  }
  \label{fig:patch-pre-filtering-and-repo-preprocessing}
\end{figure*}

With workflow overview shown in Figure~\ref{fig:patch-pre-filtering-and-repo-preprocessing}, this stage tries to identify and isolate genuine security patches from non-security-related bug fixes or refactoring.

\subsubsection{Non-Security-Related Patches Categorization}
To effectively distinguish security patches, we first establish clear categories for patches that are not primarily security-related. 
To inform this classification, we refer to previous researches on classifying noisy labels in vulnerability datasets~\cite{croftDataQualitySoftware2023,liCleanVulAutomaticFunctionLevel2024}. 
Recognizing that many identified reasons for noisy labels in prior studies exhibited considerable overlap, we consolidated and simplified these findings. 
This consolidation also clarifies the boundaries between categories, thereby alleviating the classification burden on the LLM.
This process resulted in three distinct categories of non-security-related patches:

\begin{itemize}[leftmargin=*]
\item
\textit{Testing \& Validation Updates:} This category includes patches related to testing and validation, such as debugging statements, logging, or testcases. These modifications enhance the system's efficiency or stability and are thus considered unrelated to security patches. 
For instance, in \href{https://github.com/rubiogarcia465179/TLS-GF2X/commit/97ab3c4b538840037812c8d9164d09a1f4bf11a1}{PrimeVul-506696}, the patch added a testcase for \texttt{GENERAL\_NAME\_cmp}, not addressing a security issue.

\item
\textit{Supporting \& Non-Core Improvements:} This refers to modifications outside the core logic blocks of the code, such as adding comments, changing code style, and updating configuration files. These patches primarily improve code readability or maintainability without affecting core functionality or security. 
For example, \href{https://github.com/opencv/opencv/pull/10563/files}{CVE-2018-5269} involves code refactoring where only one change is genuinely security-related, while others refactor genuine \texttt{assert} to specific \texttt{CV\_Assert}. This single commit in MegaVul generates 14 incorrect function-level labels out of 15.

\item
\textit{Defect Remediation \& Feature Upgrades:} This category includes fixes for non-security bugs or enhancements to core business logic and features. Patches fall here if they add new structures or logic to improve functionality or efficiency, or if they enhance stability without clearly addressing a security issue—making it hard to label the pre-patch code as vulnerable. This category is more complex than the previous two and often cannot be filtered by simple rules. 
For example, in \href{https://github.com/NextAlone/Nagram/commit/0bd769f842a9b40dc7f1fdf5fbb984609a883e42}{CleanVul-1138}, it optimizes Emoji matching with regular expressions to improve readability and efficiency.
\end{itemize}

\subsubsection{LLM-based Classification}

Following the definition of these categories, our methodology employs an LLM to perform a multi-step classification process.

\begin{itemize}[leftmargin=*]
\item \textit{Patch and Context Analysis:} 
First, the model analyzes the patch by thoroughly examining code diffs and all available contextual information to identify the patch's purpose, focusing on its repair strategy and technical impact, while prioritizing code-level evidence for consistency. 
While CVE descriptions are commonly included in vulnerability datasets, we intentionally exclude them from this step. We observe that CVE descriptions often emphasize the impact of the vulnerability rather than the technical details of the fix, misleading the model toward overly security-focused or generic interpretations. Instead, we focus on information that directly reflects the patch's intent and implementation. Additionally, if CVE-provided links include GitHub Pull Requests, we extract the discussion as supplementary information.
Such information, often missing from prior datasets, is vital for understanding the patch’s true purpose.

\item \textit{Security Boundary Assessment:}
The model then assesses whether the patch addresses a security boundary. This involves determining if the \textit{pre-patch} code had a condition that could, under attacker-controlled inputs or certain operational scenarios, compromise the system's intended security properties. Crucially, the patch is considered to be intended to eliminate such a condition. If both criteria are satisfied, the patch is classified as a Security Vulnerability Fix.

\item \textit{Non-Security Classification:}
If the patch does not primarily address a security boundary, the model then proceeds to classify it into one of the three categories of non-security-related patches with previous definitions provided as reference. 
Given the complexity of distinguishing ``Defect Remediation \& Feature Upgrades'' from security vulnerabilities with limited context, we prompt the model to favor a ``Security Vulnerability'' classification if it is not uncertain, prioritizing recall. Ambiguous non-security patches are further scrutinized later with more context available.

\item \textit{Confidence Scoring:} 
Finally, for each classification, the model provides a confidence score (ranging from 0.0 to 1.0) reflecting its certainty in the assigned category. Although the confidence score is not a direct measure of accuracy~\cite{openaiGPT4TechnicalReport}, it serves as a useful indicator of the model's certainty in its classification. We leverage the score to filter out low-confidence results, enhancing the overall quality of the dataset. In our experiments, we set a threshold of 0.9.

\end{itemize}

\subsubsection{Empirical Findings}%
We observe that many datasets, prioritizing rapid, large-scale construction, overlook code nuances and true patch intent. They often rely on metadata (CWE classifications, CVSS scores) or insufficient static analysis, leading to mislabeling. 
For instance, high CVSS scores may not reflect actual security vulnerabilities (e.g., \href{https://nvd.nist.gov/vuln/detail/CVE-2023-49298}{CVE-2023-49298} was a bug fix despite a 7.5 CVSS score), and \href{https://github.com/opencv/opencv/commit/be5247921da02e58aa42830c81730ef20a23af80}{all modifications} in multi-patch commits can be incorrectly flagged as security fixes, as seen with \href{https://nvd.nist.gov/vuln/detail/CVE-2018-5269}{CVE-2018-5269} in MegaVul.

Furthermore, vague CVE descriptions and commit messages often obscure the patch's true purpose. To study the influence, we trace 1,059 GitHub-sourced CVE patches, gathering richer context from PR discussions.
While commit message lengths have remained stable (29 words), PR discussions have nearly doubled in length—from 70 words pre-2020 to 144.7 post-2020—indicating increased reliance on these forums to communicate patch rationale.
Leveraging LLMs' semantic understanding, we incorporate this auxiliary information, improving classification accuracy (Section~\ref{subsubsec:impact_of_auxiliary_information}). For example, in \href{https://github.com/uclouvain/openjpeg/pull/1369}{PR} for \href{https://nvd.nist.gov/vuln/detail/cve-2022-1122}{CVE-2022-1122}, discussion enables the LLM to identify the truly security-relevant commit among three similar ones.

\subsection{Data Acquisition and Preprocessing}
\label{subsec:data-acquisition-contextual-preprocessing}

Once genuine security patches are identified, we acquire the complete source code context for these patches, as the existing datasets, like MegaVul, often provide function-level patch information instead of repository-level source code.

\subsubsection{Data Acquisition}

We use the MegaVul dataset as a starting point to extract function-level and patch metadata, then employ custom crawlers across platforms (e.g., cgit, GitHub) to retrieve corresponding full repository source code. To handle failures, our system resolves 404 errors via commit-based searches and adapts to redirects using platform-specific APIs. This approach enables successful processing of 98\% of CVEs (6,122 out of 6,269) in MegaVul.

\subsubsection{Code Property Graph (CPG) Generation}
We generate CPGs using Joern~\cite{JoernBugHunters} for the pre-patch repository to facilitate static analysis, such as retrieving the caller and callee, tracing the varaible's data-flow.
However, memory constraints sometimes prevent CPG generation for large repos like Linux, and some repositories lack specific commits, further hindering full CPG generation.
Finally, we preprocess 5,573 CVE instances.

\begin{figure*}[th!]
  \centering
  \includegraphics[width=0.85\linewidth]{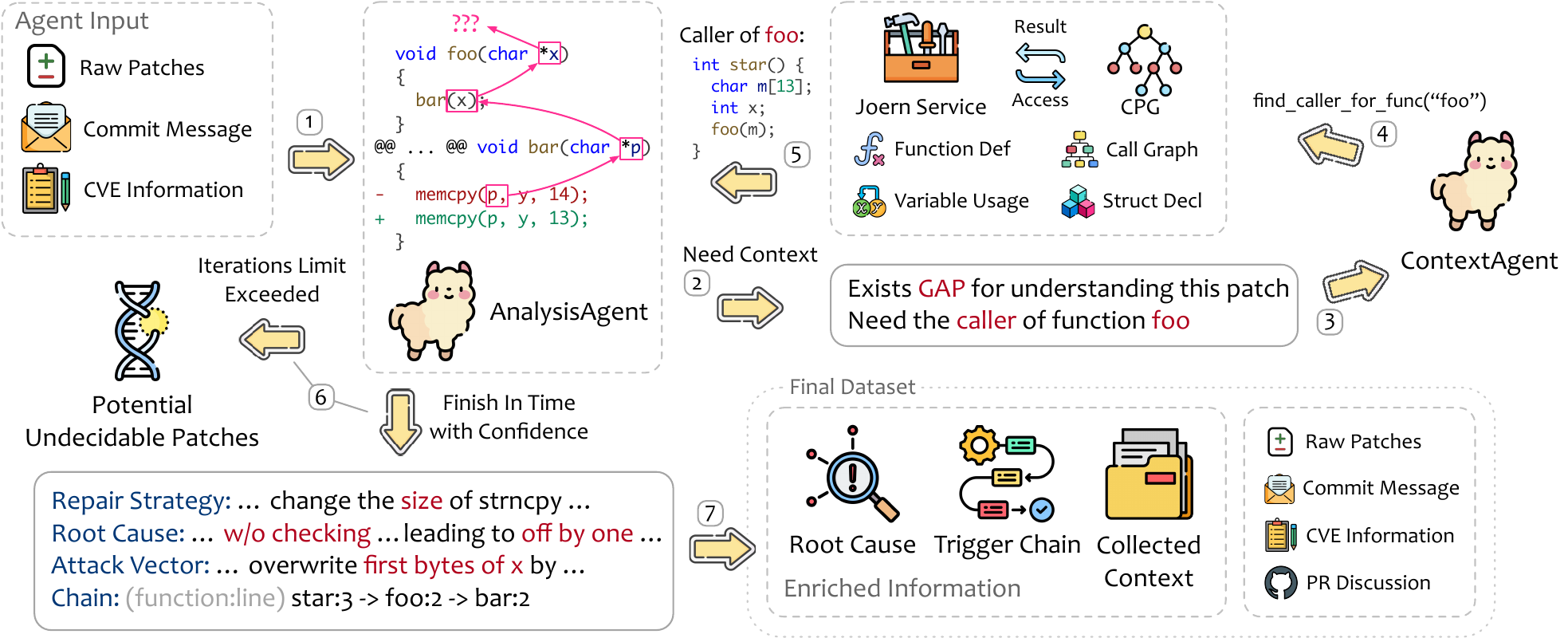}
  \caption{
    Workflow overview of iterative contextual analysis and dataset postprocessing.
  }
  \label{fig:iterative-contextual-analysis-and-dataset-postprocessing}
\end{figure*}

\subsection{Iterative Contextual Analysis}
\label{subsec:iterative-contextual-analysis}

This stage employs an LLM-driven multi-agent architecture to perform an in-depth, iterative analysis of the vulnerability, aiming to uncover its root cause and gather all semantically relevant code context. It consists of two primary agents: an AnalysisAgent responsible for understanding the collected context and a ContextAgent for context collecting, operating in a closed loop. The process is guided by a confidence score, allowing for early termination when the vulnerability is sufficiently understood or when a predefined iteration limit is reached. Fig~\ref{fig:iterative-contextual-analysis-and-dataset-postprocessing} shows the overall workflow.

\subsubsection{Root Cause Analysis and Contextual Refinement: Analysis Agent}
The AnalysisAgent commences the process with an initial assessment of the vulnerability, guided by a Zero-Assumption policy, meaning it infers nothing beyond the explicitly provided code. It then enters an iterative refinement loop, continuously seeking to extend its understanding.

\begin{itemize}[leftmargin=*]
    \item \textit{Patch Review and Initial Classification}: The agent first classifies the vulnerability into broad categories (e.g., memory-related, logic-based, or configuration issues). It then meticulously examines each hunk in the patch, explaining how it mitigates the vulnerability and citing specific file names and line numbers for each observation. This provides a foundational understanding for deeper investigation. 
    \item \textit{Trace Root Cause and Identify GAPs}: The agent attempts to trace the vulnerability's root cause by strictly following function calls and data flows within the available code (initially the patch, and then supplemented by collected context). 
    If the evidence chain breaks (i.e., a call or data flow leads outside the current scope or cannot be resolved), the agent marks it as a ``GAP'' and documents why the analysis can't continue with the available information.
    \item \textit{Formulate Context Requests}:  If critical GAPs still persist, indicating an incomplete understanding of the vulnerability's trigger chain, the agent formulates precise requests for additional information. These requests emulate an expert analyst's process, such as asking for ``the definition of function X'' or ``how variable Y is initialized,'' specifying the type of context needed (e.g., `function', `code' snippet, `caller' information, or `variable' trace). The agent avoids redundant requests and may try alternative query types if previous attempts for similar information were unfulfilled.
    \item \textit{Score Confidence}: The agent updates its confidence score based on the current completeness of the evidence chain. A high score is assigned if the full trigger chain is evident. If the chain remains incomplete, a lower score is given.
\end{itemize}

\subsubsection{Context Collection: ContextAgent}
Recognizing that not all models possess the advanced agentic capabilities of LLMs like Claude~\cite{ClaudeAgentsIntelligent} and GPT~\cite{openaiGPT4TechnicalReport,AgentsOpenAIAPI}, which can fluently invoke tools and integrate results mid-analysis~\cite{jimenezSWEbenchCanLanguage2023}, we introduce a ContextAgent.
This agent acts as a sophisticated parameter generator for static analysis tools, currently integrated with Joern. When the AnalysisAgent requests additional context, the ContextAgent translates these natural language requests into precise tool invocations.
The ContextAgent can leverage a suite of tools to gather the required information:
\begin{itemize}[leftmargin=*]
    \item Basic information retrieval: Fetching function definitions (func\_info), identifying callers of a function (caller\_info), or extracting code snippets in a specific range (code\_info).
    \item Advanced data flow and structural analysis: Tracing the definition, initialization, or usage of variables and structure members across the project (value\_info). We do not provide the full data-flow tracing capability as a tool, such as alias and points-to analysis, as it leads to timeouts when analyzing the source code in our experiments. However, we observe that the agent requests the aliased variable if it is needed, suggesting that the agent performs the alias analysis itself.
    \item Direct Joern Querying: For complex or highly specific information needs that standard tools might not cover, the ContextAgent can execute Joern queries directly (query\_info). 
\end{itemize}

By providing these fine-grained static analysis capabilities, the ContextAgent and corresponding tools empower the AnalysisAgent to obtain rich, targeted contextual information beyond simple caller-callee relationships, facilitating the understanding of complex vulnerabilities.

\subsubsection{Termination Criteria and Undecidable Patch Filtering}

This iterative process of analysis, context request, and collection continues until the agent reaches high confidence without needing further context, or a predefined iteration limit is hit. If the iteration limit is reached, \sysname discards the vulnerability, considering it a potential undecidable patch. This ensures that the \sysname either arrives at a well-supported root cause or concludes its analysis within practical limits.

\subsection{Dataset Construction and Post-processing}
\label{subsec:dataset-construction-postprocessing}

Upon completing iterative contextual analysis, all processed information is consolidated. We then construct a comprehensive dataset, indexed by CVE-ID. For each analyzed CVE and its associated patches, the dataset includes the initial patch classification, all retrieved code context (functions, data structures, call chains), the LLM-inferred root cause description, the final confidence score from \sysname, paths to relevant raw files, CWE types, and other basic information.
Built on a CVE-centric framework, this dataset boasts a low duplication rate, ensuring unique and validated entries, facilitating further research and usage for the community.

\section{Evaluation}

Our evaluation focuses on assessing the effectiveness and quality of dataset constructed by \sysname. This evaluation aims to address the following key questions:

\noindent \textbf{RQ1:} How accurately does \sysname filter non-security patches? 

\noindent \textbf{RQ2:} How effectively does \sysname extract contextual information and characterize undecidable patches?

\noindent \textbf{RQ3:} How does \sysname's gathered contextual information influence the performance of LLMs in vulnerability detection?

\noindent \textbf{RQ4:} What is the individual contribution and effectiveness of each component within the \sysname?

\subsection{RQ1: Accuracy of Non-Security-Related Fix Identification}

\subsubsection{Filtering Non-Security-Related Patches in MegaVul}
We employ DeepSeek-R1-Distill-Qwen-32B to filter 20,305 patches across 8,752 CVEs from the MegaVul dataset for patch pre-filtering and classification. The results of this initial analysis are summarized in Table \ref{tab:patch_filtering_megavul}. We select CVEs with at least one patch is Security (0.9 conf.) for the next stage of analysis, resulting in the filtering out of 28\% of the CVEs.

\begin{table}[htbp]
  \centering
  \caption{ \sysname Patch Filtering Results on MegaVul}
  \label{tab:patch_filtering_megavul}

  \scalebox{0.9}{
    \begin{threeparttable}
      \begin{tabular}{lcccc}
        \toprule
        \textbf{Category} & \textbf{\# of Java} & \textbf{Java (\%)} & \textbf{\# of C/C++} & \textbf{C/C++ (\%)}\\
        \midrule
        All patches           & 2424  & {-} & 17881 & {-} \\
        Security-related      & 1648  &  68.04 & 12315 &  68.90 \\
        Non-security-related          &  776  &  31.96 &  5568 &  31.10 \\
        \midrule[\heavyrulewidth]
        \multicolumn{5}{l}{\itshape Classification of Non-security-related}\\
        \quad Test          &   54  &   6.96 &   156 &   1.01 \\
        \quad Support         &  216  &  27.83 &   469 &   8.43 \\
        \quad Defect          &  505  &  65.14 &  4945 &  90.60 \\
        \bottomrule
      \end{tabular}
      \begin{tablenotes}[flushleft]\footnotesize
        \item Percentages are relative to the total number of patches in each language.
      \end{tablenotes}
    \end{threeparttable}
  }

\end{table}
To validate the accuracy of this critical filtering, we randomly sample 350 patches predicted as Security (0.9 conf.) and 150 predicted as non-security-related, maintaining the observed proportions. 
Three human experts independently review these samples without knowing the model's labels to establish ground truth. Our evaluation, summarized in Table \ref{tab:human_eval_matrix}, underscores the high reliability of our patch pre-filtering. For patches classified as security-relevant, the model achieved 100\% precision (350 True Positives with 0 False Positives).

{\renewcommand{\arraystretch}{1.2}
\begin{table}[htbp]
\centering
\caption{Human Evaluation of Patch Pre-filtering}
\label{tab:human_eval_matrix}
\scalebox{0.9}{
  \begin{tabular}{lcc}
  \Xhline{0.85pt}
  \textbf{Actual} \textbackslash{} \textbf{Predicted} & \textbf{Security} & \textbf{Non-security} \\
  \hline
  \textbf{Security}   & 350 (TP)                  & 19 (FN)               \\
  \textbf{Non-security}       & 0 (FP)                    & 131 (TN)               \\
  \Xhline{0.85pt}
  \end{tabular}
}
\end{table}
}
This demonstrates that when \sysname confidently identifies a security patch, it is unequivocally correct. While 19 out of 150 patches initially labeled as non-security-related were reclassified as security-related by experts, resulting in a recall of 96.2\%, these typically involved subtle like memory leaks or crashes from undefined behavior. These findings underscore the model's robust precision while highlighting areas for improvement in boundary sensitivity.

\subsubsection{Identifying Additional Non-Security-Related Patches in other dataset}

To assess \sysname's ability to uncover overlooked non-security-related patches, we randomly sample 500 samples and apply it to CleanVul\cite{liCleanVulAutomaticFunctionLevel2024} (\textit{Level 4}) and PrimeVul\cite{dingVulnerabilityDetectionCode2024} respectively.
Existing methods, like CleanVul's reliance on prompts and heuristics or PrimeVul's NVD-link-based filtering, often lead to mislabeling or omissions by incorrectly assuming patch intent or security relevance.

The results are summarized in Table~\ref{tab:other_dataset_evaluation} and manually verified by human experts for correctness. \sysname identifies 71 additional non-security-related patches in CleanVul and 106 in PrimeVul that were previously mislabeled. These include benign patches like logging code and feature additions.
We also note that while CleanVul's rule-based filter successfully filters out all `Test' patches, it shows limitations in handling `Support' and `Defect' categories, where \sysname uncovers additional misclassified instances.

{\renewcommand{\arraystretch}{1.2}
\begin{table}[htbp]
  \centering
  \caption{
    Results of filtering for other datasets.
  }
  \label{tab:other_dataset_evaluation}
  \scalebox{0.9}{
    \begin{threeparttable}
      \begin{tabular}{ccccc}
        \Xhline{0.85pt}
        \multirow{2}{*}{\textbf{Dataset}} & \multirow{2}{*}{\textbf{Count}} & \multicolumn{3}{c}{\textbf{\# of Classification}} \\ \cline{3-5}
                          &                                                           & \textbf{Test}    & \textbf{Support}    & \textbf{Defect}      \\
        \hline
        CleanVul     & 75 (69)\tnote{*}    & 0  & 4   & 71 (65)               \\
        \hline
        PrimeVul     & 116 (107)    & 4   & 6    & 106 (97)              \\
        \Xhline{0.85pt}
      \end{tabular}
      \begin{tablenotes}
        \item[*] Numbers in parentheses is reported by human experts.
      \end{tablenotes}
    \end{threeparttable}
  }
\end{table}
}

\answer {%
  \sysname achieves 100\% precision and 96.2\% recall for high-confidence security patches. Its main limitation is misclassifying around 3.8\% of borderline cases. Additionally, \sysname successfully identifies overlooked non-security-related patches in existing cleaned datasets.
}

\subsection{RQ2: Inter-Procedure and Undecidable Patches Analysis}

\subsubsection{Unveiling Inter-Procedure Vulnerability Fixes}

With Qwen3-32B as AnalysisAgent and DeepSeek-V3 as ContextAgent, 4,467 of 5,573 processed CVEs (80.15\%) were enriched with context and root causes. Table~\ref{tab:rq2_resutl_of_stage2} details average context length and top 8 CWE category distribution.

{\renewcommand{\arraystretch}{1.2}
\begin{table}[htbp]
\centering
\caption{CWE Category Distribution in \dsname}
\label{tab:rq2_resutl_of_stage2}
\scalebox{0.9}{
  \begin{tabular}{lccc}
  \Xhline{0.85pt}
  \textbf{CWE Type} & \textbf{\# of Pairs} & \textbf{pct (\%)} & \textbf{Avg. Contexts} \\
  \hline
  CWE-664 (Resource Control)     & 2,736 & 52.51 & 3.54 \\
  CWE-707 (Neutralization)       &   437 &  8.39 & 3.19 \\
  CWE-710 (Missing Functionality)&   336 &  6.45 & 3.67 \\
  CWE-703 (Exception Handling)   &   318 &  6.10 & 3.73 \\
  CWE-682 (Calculation Error)    &   302 &  5.80 & 3.70 \\
  CWE-691 (Control Flow)         &   248 &  4.76 & 2.89 \\
  CWE-284 (Access Control)       &   183 &  3.51 & 3.10 \\
  CWE-693 (Protection Failure)   &    68 &  1.31 & 2.75 \\
  Misc.                          &   582 & 11.17 & 3.16 \\
  \hline
  \textbf{Total}              & 5,210 & 100 & 3.43 \\
  \Xhline{0.85pt}
  \end{tabular}
}
\vspace{1mm}
\footnotesize{\\\textit{Note:} Total exceeds 4,467 CVEs due to multi-CWE assignments per CVE.}
\end{table}
}
A case is considered intra-procedure if its root cause lies entirely within the patched function; otherwise, it is inter-procedure. This distinction helps us assess the complexity of root cause localization for each CVE based on its reasoning scope. 
Only 493 CVEs (11\%) are intra-procedure, while the majority (89\%) require reasoning across multiple functions, showing the distributed nature of real-world vulnerabilities.

We further analyze \sysname's root cause accuracy under this classification. Table~\ref{tab:rq2_reasoning_scope_vs_validity} shows the distribution of valid and invalid root causes over 50 randomly sampled CVEs with model confidence above 0.9. Overall, 84\% of the root causes align with expert annotations. Notably, even in the more complex inter-procedure cases—which comprise 88\% of the samples—the valid rate remains high (84.1\%), demonstrating \sysname's capability to handle non-local dependencies.
A representative case is \href{https://nvd.nist.gov/vuln/detail/CVE-2022-24122}{CVE-2022-24122}, a complex kernel UAF vulnerability. While the \sysname couldn't fully resolve the root cause due to limited domain knowledge, it successfully recovered most of the trigger chain context, closely matching the developer's \href{https://www.openwall.com/lists/oss-security/2022/01/29/1}{analysis}. 
Notably, \sysname gathers context from variable and structure member usage across the codebase, beyond static call relations. This reflects a flexible, semantic exploration rather than fixed-depth call graph traversal.

\begin{table}[htbp]
\centering
\caption{Root Cause Validity by Reasoning Scope}
\label{tab:rq2_reasoning_scope_vs_validity}
\resizebox{0.90\linewidth}{!}{
\begin{tabular}{lccc}
\toprule
\textbf{Scope Type} & \textbf{\# of Cases} & \textbf{Valid Root Causes} & \textbf{Invalid Root Causes} \\
\midrule
Intra-Procedure     & 6  (12\%)  & 5  (83.3\%) & 1  (16.7\%) \\
Inter-Procedure     & 44 (88\%)  & 37 (84.1\%) & 7  (15.9\%) \\
\midrule
\textbf{Total}&      50     &     42 (84.0\%) &      8 (16.0\%)   \\
\bottomrule
\end{tabular}
}
\end{table}

\subsubsection{Identifying Undecidable Patches}

Despite \sysname's overall robustness, 1106 CVEs (19.85\%) can not be resolved within the predefined iteration limit. To understand these failures, we manually analyze 100 randomly sampled cases.

As result shown in Table~\ref{tab:undecidable_anomaly_breakdown}, 84\% are attributed to \textit{undecidable patches}—vulnerabilities that lack clear, statically verifiable signals such as discernible sources, sinks, or control/data-flow triggers (e.g., \href{https://android.googlesource.com/platform/frameworks/base/+/468651c86a8adb7aa56c708d2348e99022088af3}{CVE-2016-3838}). In such cases, \sysname often pursue speculative reasoning paths or seek non-existent evidence, resulting in exceeding the iteration limit. 

Based on this sample, we estimate that approximately 16.7\% of the 6,212 CVEs in MegaVul fall into this category. 
This indicates a non-trivial portion of the dataset that challenges automated static analysis and root cause identification. 

{\renewcommand{\arraystretch}{1.2}
\begin{table}[htbp]
\centering
\caption{Analysis of Unprocessable CVEs by \sysname}
\label{tab:undecidable_anomaly_breakdown}
\scalebox{0.9}{
  \begin{threeparttable}
  \begin{tabular}{lc}
  \Xhline{0.85pt}
  \textbf{Reason for Failure} & \textbf{Percentage (\%)} \\
  \hline
  Tool/Noise Limitations  & 16\% \\
  \textbf{Undecidable patches} & \textbf{84\%} \\
  \quad -- Runtime/High-Level Understanding & 33.3\%\tnote{*} \\
  \quad -- Complex Logic-Dependent Issues & 25.0\% \\
  \quad -- Ambiguous Defensive Programming & 21.4\% \\
  \quad -- External Knowledge/Conventions & 10.7\% \\
  \quad -- Misclassified Functional Patches & 9.5\% \\
  \Xhline{0.85pt}
  \end{tabular}
  \begin{tablenotes}
    \item[*] Percentage of Undecidable patches, same below.
  \end{tablenotes}
  \end{threeparttable}
}
\end{table}
}
Some failures also arise from the inherent complexity of certain patches. 
\sysname imposes an iteration cap, empirically set to 8, to prevent unbounded analysis. This cap, while generally effective, may lead to some manually verifiable cases exceeding the analysis limit. Determining an optimal iteration limit is an open problem, and previous works rely on empirical values~\cite{yangSWEagentAgentComputerInterfaces2024,yildizBenchmarkingLLMsLLMbased2025}. Nevertheless, \sysname performs well under its current iteration limit, balancing analytical depth with computational efficiency.
Moreover, some failures stem from misleading CVE descriptions rather than flaw complexity. For example, in \href{https://nvd.nist.gov/vuln/detail/CVE-2023-51074}{CVE-2023-51074}, a vague hint in CVE message led to an unproductive 13-step trace. Once the hint is removed, \sysname finishes the analysis of CVE in just two iterations.

\answer{%
\sysname analyzes 80.15\% of CVEs with 84\% root cause accuracy, averaging 3.43 collected contexts. 89\% of the analyzed CVEs involve inter-procedural context. This demonstrates the \sysname's capabilities of tracing vulnerabilities.
Among the unprocessable CVEs, 84\% are manually verified undecidable patches, highlighting \sysname's effectiveness in filtering undecidable patches.
}

\begin{figure*}[t!]
  \centering
  \includegraphics[width=0.85\linewidth]{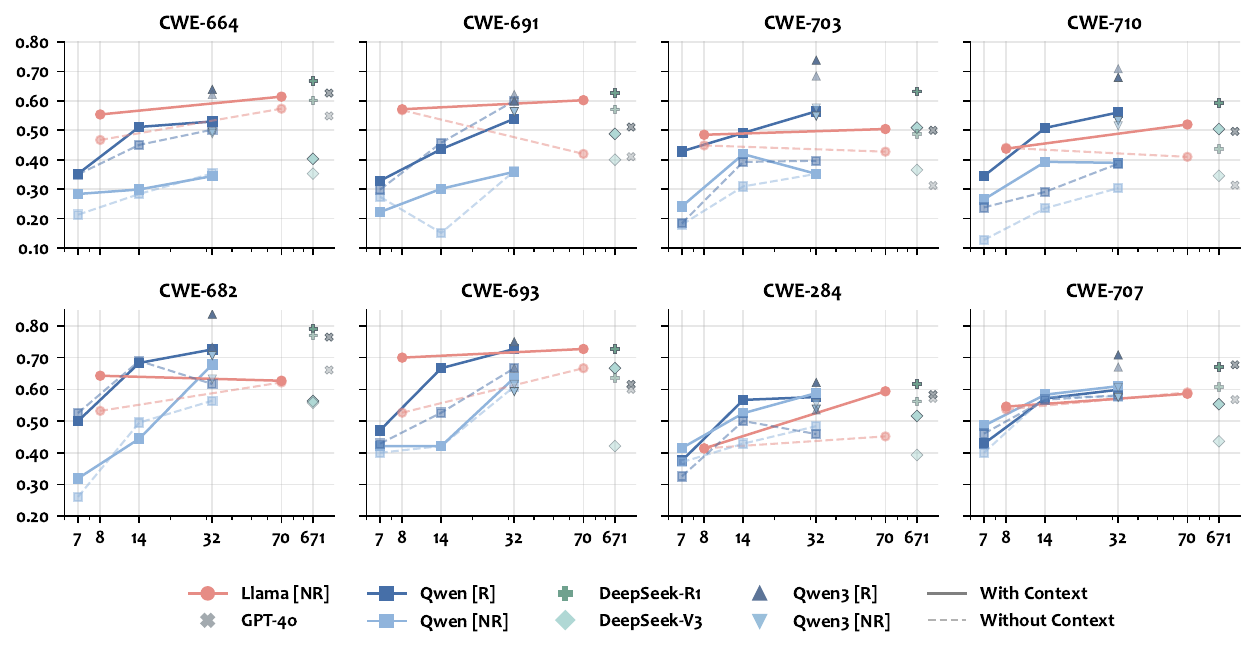}
  \caption{F1 Score of LLMs in Vulnerability Detection with and without \sysname's Context}
  \label{fig:cwe-result-f1}
\end{figure*}

\subsection{RQ3: Influence of \sysname's Context on Vulnerability Detection with LLM}

This section evaluates how contextual information gathered by \sysname impacts LLMs performance in vulnerability detection. Our objective is to quantify whether providing high-quality, relevant context enhances LLMs' ability to detect vulnerabilities and pinpoint their root causes.

\subsubsection{Experiment Setup}

\textit{Data:}
We select 1,128 CVE pairs randomly, proportionally drawn from 8 CWE categories, ensures representativeness across real-world CWE distributions. Each pair includes vulnerable and fixed function code, augmented by \sysname's gathered context. 
As a comparison, we evaluate the identical CVE pairs without context from \sysname.

\textit{Prompt:}
We design distinct prompts for two key tasks:
  (1) \textit{Vulnerability Detection (VD):} 
  We provide LLMs with task descriptions dynamically generated from CWE definitions, following~\cite{ullahLLMsCannotReliably2023}.
  Then LLMs analyze both vulnerable (pre-patch) and fixed (post-patch) code (raw, without extra annotations), performing step-by-step analysis to identify vulnerabilities and output \texttt{VUL} or \texttt{NO\_VUL}.
  (2) \textit{Root Cause Judgment (Judge):} 
  We employ an prompt containing ground truth and instruction to guide expert assessment of the root cause. The ground truth includes CVE descriptions, CWE types, patch information, and code. Judges evaluate if the LLM's identified root cause matches known information for pre-patch code and if post-patch outputs constitute false positives.

\textit{Metrics:}
We treat pre-patch code as positive (vulnerable) and post-patch code as negative (non-vulnerable), applying standard binary classification metrics as defined in Table~\ref{tab:metrics_logic}.
In pair-wise evaluation, we specifically measure the models' ability to distinguish vulnerable (pre-patch) from non-vulnerable (post-patch) versions.

\begin{table}[htbp]
\centering
\caption{Evaluation Outcomes Definition}
\label{tab:metrics_logic}

\scalebox{0.9}{
  \begin{tabular}{lcccccc}
  \toprule
  & \multicolumn{3}{c}{\textbf{Pre-patch Code}} & \multicolumn{3}{c}{\textbf{Post-patch Code}} \\
  \midrule
  \textbf{VD (Pred)} & T & F & T & T & T & F \\
  \textbf{Judge}     & T & - & F & F & T & - \\
  \textbf{Result}    & TP & FN & FN & TN & FP & TN \\
  \bottomrule
  \end{tabular}
}
\end{table}

\textit{Models:}
To assess the influence of context across model architectures and scales, we select 13 diverse LLMs from Deepseek, Qwen, Meta and OpenAI. We denote the reasoning model as \texttt{[R]} and the non-reasoning model as \texttt{[NR]} for each model in the suffix if available.

\subsubsection{Result}
\begin{figure}[t!]
  \centering
  \includegraphics[width=0.99\linewidth]{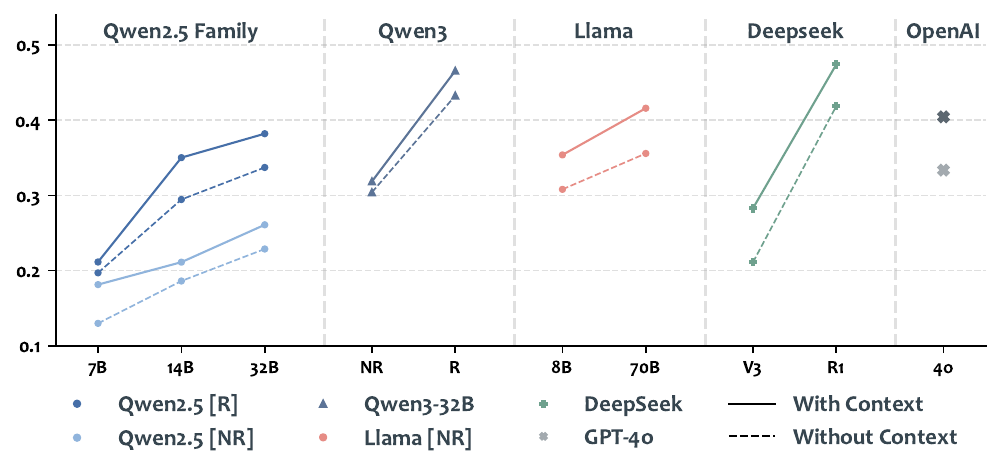} 
  \caption{Paired Accuracy with and without \sysname's Context}
  \label{fig:cwe-pair-accuracy}
\end{figure}
Our evaluation of 1,128 CVE pairs (Figure~\ref{fig:cwe-result-f1}) revealed consistent patterns across models. Specifically for CWE-664, the prevalent vulnerability, models without context typically achieved an F1 of 0.5 (max 0.62 for \texttt{Qwen3 [R]}). Crucially, adding context improved performance by 3\textasciitilde 8\%, with DeepSeek-R1 seeing the biggest jump to 0.67.
\sysname consistently improves F1-scores by 3\%\textasciitilde 15\% across various CWE types for comparable models, due to its concise, high-quality context that aids in quickly identifying key variables and control flow. Larger models further benefit, with performance gains extending to complex CWEs like CWE-284 (Access Control) and CWE-707 (Neutralization).

Figure~\ref{fig:cwe-pair-accuracy} presents the results from our paired detection experiments across all detected pairs. We observe that all models demonstrated an improvement in paired detection rates ranging from 1.5\textasciitilde7.2\% when compared to evaluations without context. Among these, DeepSeekR1 achieves a remarkable paired detection rate of 47\% across all vulnerability pairs.

\answer{\sysname's context improves LLMs’ performance on vulnerability detection tasks, with 3\textasciitilde15\% gains in F1-score and 1.5\textasciitilde7.2\% in paired detection accuracy, demonstrating the value of \sysname's collected context.
}

\subsection{RQ4: Individual Component Contribution of \sysname}

\subsubsection{Impact of Auxiliary Information on Patch Classification}
\label{subsubsec:impact_of_auxiliary_information}
To study the impact of our enriched auxiliary information, including PR information and detailed comments, on patch classification,we evaluate only using commit messages and raw patch code from 500 random patches, as CleanVul and PrimeVul do. It shows that LLM disagrees with previous classifications in 86 instances.

However, human assessment confirmed that classifications guided by auxiliary information, aligned more closely with human intuition. This comprehensive context improved the LLM's ability to discern patch intent and security relevance. Specifically, 78 of these 86 instances are re-evaluated and confirmed as correctly classified by the LLM when it is provided with the enriched auxiliary information.

\subsubsection{Impact of Specialized Tools on Context Extraction}
We further assess the contribution of \sysname's fine-grained static analysis tools, such as Value Trace, arbitrary code snippet queries, and struct queries, by removing the access to them, permitting only basic caller and function info tools. 
From our successfully processed dataset, we select 100 CVEs for which experts confirm accurate root cause identification and correct context collection.
The results for Agent context construction performance under these limitations are summarized below:

{\renewcommand{\arraystretch}{1.1}
\begin{table}[htbp]
\centering
\caption{
  Impact of Tool Restriction 
}
\label{tab:tool_limitation_impact}
\begin{tabular}{lc}
\toprule
\textbf{Outcome Category} & \textbf{Count (out of 100)} \\
\midrule
Successfully Processed & 54 \\
Unprocessable (Hit Iteration Limit) & 46 \\
\bottomrule
\end{tabular}
\end{table}
}

As Table~\ref{tab:tool_limitation_impact} shows, 
without its advanced tools, \sysname often asked for more context but still failed over 46\% of the time on the test set because it hit its maximum request limit. This suggests that without its specialized tools for data and code retrieval, \sysname struggles to trace complex vulnerabilities and frequently reaches its operational limits.

\answer{%
  \sysname's effectiveness relies on its core components: enriched ground truth improved patch classification accuracy by 15.6\%, and specialized tools were crucial for context extraction, preventing over 46\% of resolution failures. Both are indispensable for \sysname's performance.
}

\section{Related Work}

\textbf{Vulnerability Dataset Construction.}
The evolution of vulnerability datasets has been marked by a shift from prioritizing quantity to emphasizing quality and contextual richness. 
Early large-scale efforts in vulnerability dataset construction vary. BigVul~\cite{fanBigVulCodeVulnerabilityDataset2020} amasses C/C++ function-level vulnerabilities by systematically mining public databases (CVE~\cite{CVECommonVulnerabilities}, NVD~\cite{NVDNatuionalVulnerability}) and automatically retrieving code, emphasizing automated processes. CrossVul~\cite{nikitopoulosCrossVulCrosslanguageVulnerability2021} also mines such sources, offering multi-language coverage at file-level granularity. CVEFixes~\cite{bhandariCVEfixesAutomatedCollection2021} provides structured, multi-level data by retrieving information associated with CVE records. In contrast, Devign~\cite{zhouDevignEffectiveVulnerability2019} takes a distinct early approach by prioritizing high-quality manual labeling.
Subsequent datasets continue to evolve: DiverseVul~\cite{chenDiverseVulNewVulnerable2023} aims to enhance realism and diversity by improving automated label accuracy for C/C++ data and broadening project/CWE diversity. MegaVul~\cite{niMegaVulVulnerabilityDataset2024} pursues greater scale and quality, featuring enhanced parsing via Tree-sitter~\cite{TreesitterTreesitter2025}, richer code representations like PDGs, and multi-language support.

\textbf{Deep Learning and LLM-based Vulnerability Detection.}
Research in deep learning-based vulnerability detection has transitioned from specialized, custom-designed models to leveraging the broad capabilities of Large Language Models (LLMs).
Early approaches vary: ReVeal~\cite{chakrabortyDeepLearningBased2020} utilizes Graph Neural Networks (GNNs) on code property graphs; VulCNN~\cite{wuVulCNNImageinspiredScalable2022} innovatively treats code as images for CNN-based analysis; and LineVul~\cite{fuLineVulTransformerbasedLineLevel2022} employs Transformers like CodeBERT for fine-grained line-level prediction.
While the advent of LLMs has shown considerable potential for vulnerability identification, comprehensive benchmarks such as VulnLLMEval~\cite{zibaeiradVulnLLMEvalFrameworkEvaluating2024}, VulDetectBench~\cite{liuVulDetectBenchEvaluatingDeep2024}, VulBench~\cite{gaoHowFarHave2023}, JitVul~\cite{yildizBenchmarkingLLMsLLMbased2025} have also highlighted their current limitations. These limitations primarily lie in achieving deep semantic understanding of vulnerabilities and performing precise, fine-grained localization of vulnerable code.

\section{Limitations \& Discussion}

Our work, while demonstrating promising results, has several limitations and some potential future work.
\subsubsection{Reliance on Manual Tool Implementation}
\sysname's effectiveness relies on its integrated static analysis tools. 
Current tool limitations, like unhandled corner cases or missing features (e.g., tracing unassigned variable paths), can impede context acquisition and lead to analysis failures. 
Although LLM can generate Joern queries directly, the LLM's unfamiliarity with Joern's syntax yield a low success rate (around 12\% non-empty results) for such ad-hoc queries. 
Developing more robust, pre-defined tools that encapsulate Joern's capabilities is a more viable near-term solution.

\subsubsection{Classification of Undecidable patches}
Our proposed classification of undecidable patches may not cover all types. 
The current approach uses agent understanding to filter these patches, but agent limitations, not definitive undecidability, lead to some discards. 
Future work can refine the definition and identification methods for undecidable patches. 
Besides, identifying patches that require external knowledge remains a challenge. Effectively filtering and using this subset for model learning is an important future research direction.

\subsubsection{Bridging the Gap to Practical Vulnerability Detection}
Our experiments show that \sysname's context yields positive performance improvements for LLMs in vulnerability detection.
However, a notable gap still persists between current LLM capabilities and practical application.
This paper's dataset filtering and construction methodology aims to produce a cleaner training dataset.
Exporing training models on such a noise-reduced dataset is crucial future work to enhance model performance and usability.

\section{Conclusion}

In this paper, we identify and mitigate several prevalent types of noisy patches in vulnerability datasets that degrade model performance. While prior work mainly focus on noise caused by \textit{semantic mislabeling} and \textit{inter-procedure ambiguity}, we introduce and formalize a novel noise category: \textit{undecidable patches}. 
To address these noisy patches, we propose \sysname, an LLM-powered multi-agent framework that emulates human expert analysis. \sysname classifies patches, performs iterative contextual analysis, and filters undecidable patches. 
Our evaluations show that \sysname significantly improves dataset quality by filtering mislabeled instances that constituted 31\% of the MegaVul dataset and filtering undecidable patches that represented another 16\% of this dataset.
The improved datasets boost LLM-based vulnerability detection performance by 15\%. 
Both \sysname and the \dsname dataset are open-sourced to facilitate further research in vulnerability detection.

%% file: comparison-table.tex
{\renewcommand{\arraystretch}{1.2}

    \begin{table}[t!]
        \centering
        \caption{Comparison of works for handling noisy patches.}
        \resizebox{0.98\linewidth}{!}{%
        \label{tab:comparison-table}
    
        \begin{threeparttable}
            \begin{tabular}{l c c c c c c c}
                \Xhline{0.85pt} 
                \textbf{\makecell{Research Works}}                                                                                                                                                                                                                                                                                                  & \textbf{\makecell{Correct \\ Semantic Labeling}} & \textbf{\makecell{Handling \\ Inter-procedural}} & \textbf{\makecell{Pruning \\ Undecidable Patches}} \\
                \makecell[l]{MegaVul~\cite{niMegaVulVulnerabilityDataset2024}, DiverseVul~\cite{chenDiverseVulNewVulnerable2023} \\ CVEFixes~\cite{bhandariCVEfixesAutomatedCollection2021}, BigVul~\cite{fanBigVulCodeVulnerabilityDataset2020} \\ ReVeal~\cite{chakrabortyDeepLearningBased2020}, CrossVul~\cite{liCrossdomainVulnerabilityDetection2023}} & \xmark                                           & \xmark                                           & \xmark                                             \\
                \rowcolor{gray!10} ReposVul~\cite{wangReposVulRepositoryLevelHighQuality2024}, VulBench~\cite{gaoHowFarHave2023}                                                                                                                                                                                                                             & \tmark                                           & \tmark                                           & \xmark                                             \\
                \makecell[l]{Devign~\cite{zhouDevignEffectiveVulnerability2019}, PrimeVul~\cite{dingVulnerabilityDetectionCode2024} \\ CleanVul~\cite{liCleanVulAutomaticFunctionLevel2024}}                                                                                                                                                                 & \tmark                                           & \xmark                                           & \xmark                                             \\
                \rowcolor{gray!10} CORRECT~\cite{liEverythingYouWanted2025}, Vultrigger~\cite{liEffectivenessFunctionLevelVulnerability2024}                                                                                                                                                                                                                                                                                  & \xmark                                           & \tmark                                           & \xmark                                             \\
                \textbf{\sysname (Ours)}                                                                                                                                                                                                                                                                                                                                & \cmark                                           & \cmark                                           & \cmark                                             \\
                \Xhline{0.85pt} 
            \end{tabular}%
    
            \begin{tablenotes}
                [flushleft]\footnotesize \item Symbol: \cmark \ (full support), \tmark \ (partial), \xmark \ (none).
            \end{tablenotes}
        \end{threeparttable}
    
        }%
        \label{tab:full-comparison}
    \end{table}
}